\documentclass{ws-procs975x65}

\begin{document}

\title{VIOLATION OF CAUSALITY IN $\mathbf{f(R)}$ GRAVITY}
\vspace{-2mm}
\author{M. J. REBOU\c{C}AS}
\address{Centro Brasileiro de Pesquisas F\'{\i}sicas, 22290-180 Rio de Janeiro--RJ,  Brazil}
\vspace{-2mm}
\author{J. SANTOS}
\address{Universidade Federal do Rio G. do Norte, 59072-970 Natal -- RN, Brazil} 


\vspace{-4mm}
\begin{abstract}
We examine the question as to whether the $f(R)$ gravity theories, in both metric
and in Palatini formalisms, permit space-times in which the causality is violated.
We show that the field equations of these $f(R)$ gravity theories admit
solutions with violation of causality for a physically well-motivated perfect-fluid
matter content.
\end{abstract}


\bodymatter

\vspace{-2mm}
\section{Introduction} \label{Intro}
Generalization of the Einstein-Hilbert Lagrangian by replacing the scalar curvature
$R$ by a general function $f(R)$, gives rise to theories known as $f(R)$ gravity.
These theories have recently received considerable attention (see,
e.g., Refs.~\refcite{papers} and references therein) motivated mainly
by the fact that they can explain the observed accelerating
late expansion of the universe with no dark energy component.
They also offer a much richer framework than General Relativity (GR) in the
sense that $f(R)$ theories has two different formulations known as metric
and Palatini formalisms, which give rise to very different equations of motion.
In the metric formulation it is assumed that the metric and connections are
compatible, while in the Palatini formulation the equations of motion are
obtained by considering the metric and connection as independent fields
in the variation of the action.
Much efforts from the theoretical viewpoint have been developed so far
in order to clarify inherent subtleties of these theories, including
general principles such as the so-called energy conditions\cite{energy_conditions}.
Another interesting issue is the question as to whether these theories
permit space-time solutions in which causality is violated.
It is well-known that in GR there are solutions to the field equations
that possess causal anomalies in the form of closed time-like curves.
The solution found by G\"odel\cite{Godel49} is the best known example.
In this context, we have recently examined the violation of causality in both
formulations of $f(R)$ gravity\cite{Reboucas}. Here we extend the previous
works by presenting an unified view of causality problem in both versions
$f(R)$ gravity.

\section{Metric and the Palatini Approaches} \label{Metric}
The action that defines an $f(R)$ gravity is given by
\begin{equation}  \label{action}
 S=\int d^4x\sqrt{-g}\,\,\left[\,\frac{f(R)}{2\kappa^2} + \mathcal{L}_m
 \right] \,,
\end{equation}
where $\kappa^2\equiv 8\pi G$, $g$ is the determinant of the metric
$g_{\mu\nu}$ and $\mathcal{L}_m$ is the Lagrangian density for the matter fields.
Varying this action with respect to the metric we obtain the field equations
\begin{equation}  \label{field_eq}
f_RR_{\mu\nu} - \frac{f}{2}g_{\mu\nu} - \left(\nabla_{\mu}\nabla_{\nu}-
g_{\mu\nu}\,\Box \,\right)f_R = \kappa^2T_{\mu\nu}\,,
\end{equation}
where $f_R\equiv df/dR$, $\Box = g^{\alpha \beta}\,\nabla_{\alpha}\nabla_{\beta}\,$,
and $T_{\mu\nu}$ is the energy-momentum tensor.

In the Palatini formulation we treat the metric and the connection as
independent fields, and variation of the action (\ref{action}) gives
the field equations
\begin{equation}
\label{field_eq-Pal}
f_RR_{(\mu\nu)} - \frac{f}{2}g_{\mu\nu}  = \kappa^2T_{\mu\nu}\,.
\end{equation}
Here the Ricci tensor $R_{\mu\nu}$ must be calculated using the connections
field given by
\begin{equation}  \label{Gamma}
\Gamma_{\mu\nu}^{\rho} = \left\{^{\rho}_{\mu\nu}\right\}
+ \frac{1}{2f_R}\left( \delta^{\rho}_{\mu}\partial_{\nu} +
\delta^{\rho}_{\nu}\partial_{\mu} - g_{\mu\nu}g^{\rho\sigma}\partial_{\sigma} \right)f_R\,,
\end{equation}
where $\left\{^{\rho}_{\mu\nu}\right\}$ are the Levi-Civita connections of the metric
$g_{\mu\nu}$.

\vspace{-2mm}
\section{G\"{o}del-type Geometry} \label{Godel}
The homogeneous G\"odel-type  metric in cylindrical
coordinates $(r, \phi, z)$ is given by\cite{Reb_Tiomno}
\begin{equation}  \label{G-type_metric}
ds^2=dt^2 +2\,H(r)\, dt\,d\phi -dr^2 -G(r)\,d\phi^2 -dz^2 \,,
\end{equation}
where $G(r)\equiv D^2 - H^2$, $H(r) = (4\omega/m^2)\,\sinh^2(mr/2)$ and $D(r) = \sinh(mr)/m$,
with $\omega$ and $m$ being constant parameters such that $\omega^2 > 0$ and
$-\infty\leq m^2\leq +\infty$.
The G\"odel solution\cite{Godel49} is a particular case in which $m^2= 2 \omega^2$.
The existence of closed time-like curves of G\"odel-type depends on the behavior
of the function $G(r)$. If $G(r) < 0$ for a certain range of $r$ ($r_1 < r < r_2$, say)
\emph{G\"odel circles} defined by  $t, z, r = \text{const}$ are closed time-like curves.
The causality features of the  G\"odel-type spacetimes thus depend on the two independent
parameters $m$ and $\omega$. For $m=0$ there is a critical radius, defined by $\omega r_c = 1$,
such that for all $r>r_c$ there are noncausal G\"odel circles. For $0 < m^2 < 4\omega^2$
noncausal G\"odel circles occur for $r>r_c$ such that $\sinh^2(mr_c/2) = 1/(4\omega^2/m^2 - 1)$.
Clearly this type of violation
of causality is not of trivial topological nature, which are
obtained by topological identification~\cite{Topology}.

\vspace{-4mm}
\section{Non-causality in $\mathbf{f(R)}$ Gravity}
We can simplify our calculations choosing the tetrad basis
$(\theta^0, \theta^1, \theta^2, \theta^3) = (dt + H(r)d\phi,\,dr, \,D(r)d\phi,\,dz)$,
relative to which the field equations in both formalism [Eq.~(\ref{field_eq}) and
Eq.~(\ref{field_eq-Pal})] reduce to
\begin{equation}  \label{G_AB}
f_R\,G_{AB} = \kappa^2\,T_{AB} - \frac{1}{2}\left( \kappa^2T + f \right)\eta_{AB}\,,
\end{equation}
where $T$ is the trace of energy-momentum tensor $T_{AB}$ and we have taken into
account that for G\"odel-type metrics the Ricci scalar is constant $R = 2 (m^2 - \omega^2)$.
An important ingredient of the causality problem in G\"{o}del-type universes is
the matter source, which consider a perfect fluid of density $\rho$ and pressure $p$.
Thus, the field equations (\ref{G_AB}) give us
\begin{eqnarray}
2(3\omega^2 - m^2)f_R + f &=& \kappa^2\,(\rho + 3p)\,,\label{1st-eq}\\
2\omega^2f_R - f &=& \kappa^2\,(\rho - p)\,, \label{2nd-eq} \\
2(m^2 - \omega^2)f_R - f &=& \kappa^2\,( \rho - p)\,.\label{3rd-eq}
\end{eqnarray}
Equations (\ref{2nd-eq}) and (\ref{3rd-eq}) give $(m^2 - 2\omega^2)f_R = 0$.
Hence, for $f(R)$ theories that satisfy the condition $f_R> 0$, we must have
$m^2 = 2 \omega^2$ which defines the G\"odel metric.
The above field equations are then rewritten as
\begin{equation}
\kappa^2 p = f/2 \qquad \mbox{and} \qquad \kappa^2 \rho = m^2 f_R - f/2\,,
\end{equation}
where $f$ and $f_R$ are valuated at $R=m^2$. This result can be seen as a extension of
Bampi and Zordan\cite{BampiZordan78} GR result to the context of $f(R)$ gravity in
both formulation, in the sense that for arbitrary $\rho$ and $p$ (with $p \neq -\rho$)
every perfect fluid G\"odel-type solution of $f(R)$ gravity, which satisfies the
condition $f_R>0$, is necessarily isometric to the G\"odel geometry.
Concerning the causality features of these solutions we first note that
since they  are isometric to G\"odel geometry they exhibit noncausal G\"odel
circles for $r>r_c$.
On the other hand, taking into account the above results
we have that, in the framework of $f(R)$ gravity,  $r_c$ is given by
$r_c = 2 \, \sinh^{-1} (1)\, \sqrt{2f_R/(2\kappa^2\rho + f )}$,
making apparent that the critical radius depends on both the
gravity theory and the matter content.

\vspace{-4mm}
\section*{Acknowledgments}
M.J.R. acknowledges the support of FAPERJ under a CNE grant.
M.J.R. and J.S. thanks CNPq for the grant under which this work
was carried out. J.S. also thanks financial support from Instituto Nacional de Estudos do
Espa\c{c}o (INEspa\c{c}o), Funda\c{c}\~ao de Amparo a Pesquisa do RN (FAPERN).
We are grateful to A.F.F. Teixeira for indicating omissions and misprints.

\end{document}